January 2020

# AWARENESS AND USAGE OF SWAYAM COURSES AMONG LIBRARY AND INFORMATION SCIENCE STUDENTS: A SURVEY

A SUBAVEERAPANDIYAN Mr
*Library Trainee, IIM Nagpur*, subaveerapandiyan@gmail.com

Fakrudhin Ali Ahamed H Mr
*Research Scholar, Central University of Tamil Nadu, Thiruvarur*, faa.ctr@gmail.com

# AWARENESS AND USAGE OF SWAYAM COURSES AMONG LIBRARY AND INFORMATION SCIENCE STUDENTS: A SURVEY

SUBAVEERAPANDIYAN A* and H. FAKRUDHIN ALI AHAMED**

*Library Trainee, IIM Nagpur, email: subaveerapandiyan@gmail.com
** Research Scholar, Department of Library & Information Science, Central University of Tamil Nadu, Thiruvarur, e-mail: faa.ctr@gmail.com

**Abstract**

The Government of India initiated SWAYAM (Study Webs of Active-learning for Young Aspiring Minds), where the objective of the programme is to take the best teaching and learning resources to all with no costs. To find out the awareness and usage of SWAYAM courses among Library and Information science students, survey method of research used. To collect data from selective Universities students a structured questionnaire was prepared. The results show that most of the responses received from Annamalai University, majority respondents are aware of SWAYAM course. Respondents are aware of these courses through their teachers. 74.73% of respondents replied that their University providing orientation programme on SWAYAM. Half of the respondents prefer Word file format for submitting their assignment. More than 82% of respondents are spending 1-3 hours for SWAYAM course. Half of the respondents are agreed SWAYAM courses help to gain new knowledge and supports life-long learning. Further, the result reveals that the numbers of students continue their enrolled courses and only less number were discontinued due to some reasons.

**Keywords**: Swayam, MOOCs, OER, Library & Information Science, Open Course Ware.

## 1. Introduction

The commencement of the open access in education was marked by the OpenCourseWare (OCW) initiative of the Massachusetts Institute of Technology (MIT) which uploaded most of their course materials on the Web in 2001, thus making them accessible worldwide for free of cost. This was set as an example which was followed by several world-renowned universities which, extended their influence both within the academic community and among people who desire to learn. The concept of Open Educational Resources was mentioned for the first time in 2002 at the UNESCO Forum on Open Courseware for Higher Education, emphasizing the idea of free sharing of knowledge and digital teaching, learning and research materials (Butcher, 2011; Poposki, 2010). Open Educational Resources (OER) include any educational and research resources as well as curriculum maps, course materials, entire and parts of E-courses, lessons plan, learning materials, textbooks, audio and video records, simulations, experiments, multimedia content, applications, games and any other materials that have been designed for use in teaching, learning and researching that are openly available for use without an accompanying need to pay fees (Butcher, 2011; Groom, 2013). "Access to information and knowledge are the fundamental right for every human", but this is not achieved forever without restrictions. India is home to the oldest education system (Sharma & Sharma, 2012) and has one of the world's largest higher education systems (World Bank, 2007), paradoxically it also has the lowest literacy rate (UNESCO, 2014). Open educational resources (OERs) have the potential to change the educational landscape on a global scale, particularly that of developing nations (Richter & McPherson, 2012), by offering access to high quality educational resources without cost (Moore, 2015) and despite a poorly developed educational infrastructure (Richter & McPherson, 2012).

2. About SWAYAM

SWAYAM (Study Webs of Active-learning for Young Aspiring Minds) programme initiated by the Government of India. The objective of the programme is to take the best teaching-learning resources to all. Currently, it is in developing stage covering various courses includes school, certificate courses, diploma, under-graduate, post-graduate courses. In future, it would be introduced in engineering, law and also in other professional courses. SWAYAM is designed to achieve the three cardinal principles of Education Policy viz., access, equity and quality. More than 1,000 specially selected faculty and teachers across the country have contributed in preparing these courses. The courses hosted on SWAYAM are in 4 quadrants – (1) video lecture, (2) specially prepared reading material that can be downloaded/printed (3) self-assessment tests through tests and quizzes and (4) an online discussion forum for clearing the doubts. In order to ensure best quality content are created and distributed, nine National Coordinators have been chosen: They are AICTE for self-paced and international courses, NPTEL for engineering, UGC for non-technical post-graduation education, CEC for under-graduate education, NCERT & NIOS for school education, IGNOU for out of the school students, IIMB for management studies and NITTTR for Teacher Training programme[1].

## 3. Related Studies

Numerous studies carried out on OER and Massive open online courses (MOOCs) some of them discussed here. Allen and Seaman (2013) in their report student retention in online programmes are particularly relevant to the discussion of student satisfaction with their online experience. Michele T. Cole, Daniel J. Shelley, and Louis B. Swart (2014) in their study found that online instruction as moderately satisfactory, with hybrid or partially online courses rated as somewhat more satisfactory than fully online courses. U.S. Department of Education, (2010) study found that students who took all or part of their course online performed better, on average than those taking the same course through traditional face-to-face instruction. ElkeHöfler and Claudia Zimmermann, (2017) their study found that MOOCs are boasting considerable participant numbers, but also suffer from declining participant activity and low completion rates. Nikola Stevanovic (2014) the paper based on the view that the avid number of students enrolled in online courses does not necessarily finish them and earn the certificate. Specialized web sites such as Coursera, EdX and Udacity are getting more members each day, but the number of people who attend courses is not getting higher. Paul Diver and Ignacio Martinez (2015) Massive open online courses (MOOCs) offer many opportunities for research into several topics related to pedagogical methods and student incentives. They considered experiments are needed that will decide how students change behaviour when they are offered Certificates of different values. Shamprasad M. Pujar and Prahalad G. Tadasad (2016) their paper aims to explore how this new model of education can bring opportunities to LIS schools to overcome such constraints as lack of teachers, variable skills levels, a paucity of funds and limited infrastructure, all of which can be significant barriers to the effective delivery of LIS education. Neil Smith, et al. (2017) the paper is to present a comparison of two ways of developing and delivering massive open online courses (MOOCs). The study found MOOCs on existing large platforms can reach thousands of people, but constrain pedagogical choice. Self-made MOOCs have smaller audiences but can target them more effectively. Form the above studies it is found that nobody is undertaken to study the Library and Information Science students awareness and Usage of SWAYAM courses. Hence, the researcher undertook this study.

---

[1] https://swayam.gov.in/about

**Objectives of the Study**

- To find out the awareness of SWAYAM Course.
- To find out the actual time spent on learning in SWAYAM Course.
- To find out the problems faced by students while pursuing the courses.
- To find out the learning outcome of the student through SWAYAM Courses.

**4. Methodology**

In this study, a survey method was used. For data collection structured questionnaire was designed. The sample of 300 students was selected using proportionate stratified random sampling in selected Universities. The data was collected during the month of February-April 2019. For collecting data the investigator personally visited each University and distributed 300 questionnaires among the students. Out of which, 186 (62%) questionnaires dully filled in were received back shown in table 1. To analyze the collected data MS-Excel used. The descriptive statistics including percentages and graphic representations are used to provide a general picture of the awareness and usage of SWAYAM Course among the Library and Information Science students.

**Table 1 Response rate of the Students**

| Questionnaires | Number | Percentage |
|---|---|---|
| Received | 186 | 62 |
| Not Received | 114 | 38 |
| Total | 300 | 100 |

## 5. Limitations of the Study

The present study is limited to selective University Students. Due to lack of time and money, the sample survey was limited to the library and information science students of selected Universities.

## 6. Analysis and Discussion

### 6.1 Demographic Data

Table 2 shows the University wise distribution of the responses received from the students of various Universities. The analysis indicates that out of 186 respondents 41.94% of the responses were participants from Annamalai University and 13.97% responses were received from the Central University of Tamil Nadu. Among the respondents, 11.29% responses were from Periyar University and 10.22% of them Bharathidasan University. About 12.91% and 9.67% of the respondents are from the University of Madras and Pondicherry University. It can be seen that 87.63% of the respondents belong to the age group of 18-24 years followed by 9.13% of the respondents belong to the age group of 25-30 years. Only 2.16% and 1.08% respondents are from the age group of 31-44 years and above 40 years respectively. Majority of 67.74% of the respondents are Male and the remaining 32.26% are female, no one represents transgender.

Table 2: University, Age and Gender wise Responses.

| Name of University | Respondents | Age | | | | Gender | | |
|---|---|---|---|---|---|---|---|---|
| | | 18-24 | 25-30 | 31-40 | >40 | Male | Female | Trans Gender |
| Annamalai University | 78 (41.9) | 68 | 8 | 2 | 0 | 58 | 20 | 0 |
| Central University of Tamil Nadu | 26 (13.97) | 22 | 2 | 0 | 2 | 12 | 14 | 0 |
| Periyar University | 21(11.29) | 18 | 2 | 1 | 0 | 16 | 5 | 0 |
| Bharathidasan University | 19(10.22) | 16 | 2 | 1 | 0 | 12 | 7 | 0 |
| University of Madras | 24 (12.91) | 23 | 1 | 0 | 0 | 16 | 8 | 0 |
| Pondicherry University | 18 (9.67) | 16 | 2 | 0 | 0 | 12 | 6 | 0 |
| Total | 186 | 163 (87.63) | 17 (9.13) | 4 (2.16) | 2 (1.08) | 126 (67.74) | 60 (32.26) | 0 |

(Within the parenthesis indicates percentage)

**6.2 Awareness of SWAYAM**

From Table 3 it shows that 83.33% of the respondents are aware of SWAYAM courses remaining 16.67% are unaware.

Figure 1: Awareness of SWAYAM

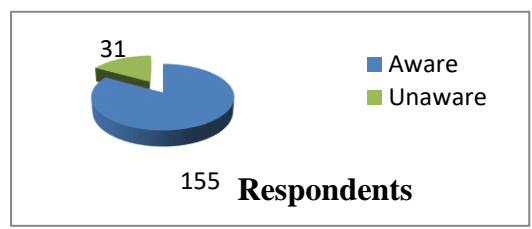

Table 3: Awareness of SWAYAM

| SWAYAM Courses | Respondents | Percentage |
|---|---|---|
| Aware | 155 | 83.33 |
| Unaware | 31 | 16.67 |
| Total | 186 | 100 |

**6.3 Source of awareness of SWAYAM**

From table 4 indicates the sources from which the respondents came to know about the SWAYAM and learnt about its use. The analysis shows that the majority of 41.29% of the respondents known about the SWAYAM from teachers and 22.58% from friends. About 19.36% of them came to know about SWAYAM from University programs like conference/seminar/workshop, 10.32% from social media, 3.87% from other sources and only 2.58% from library networks.

Table 4: Source of awareness of SWAYAM

| Sources | Respondents | Percentage |
|---|---|---|
| Teachers | 64 | 41.29 |
| Friends | 35 | 22.58 |
| University Program | 30 | 19.36 |
| Social Media | 16 | 10.32 |
| Others | 6 | 3.87 |
| Library Networks | 4 | 2.58 |
| Total | 155 | 100 |

**6.4 University organized an orientation program about SWAYAM**

The question has been asked whether University organized orientation program regarding SWAYM, the analysis shows that a good number of respondents 74.73% replied positively remaining 25.27% replied negatively.

**Figure 2 and Table 5: University Organized orientation program about SWAYAM**

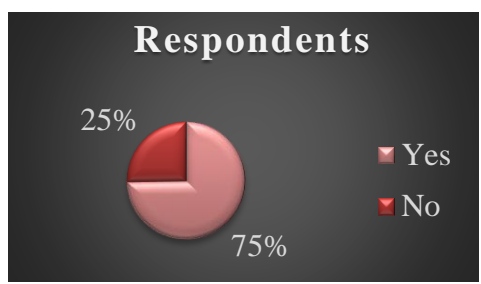

| Orientation for SWAYAM | Respondents | Percentage |
|---|---|---|
| Yes | 139 | 74.73 |
| No | 47 | 25.27 |
| Total | 186 | 100 |

**6.5 Organizing frequency of orientation program on SWAYAM**

The orientation programme is the most of academic activity. From table 6 the analysis found that 74.1% of the respondents are responding university conducted orientation program annually, followed by 23.74% respondents are replied half-yearly, only a few respond monthly and nil response for weekly.

Table 6: Organizing the frequency of orientation program on SWAYAM

| Frequency | Respondents | Percentage |
|---|---|---|
| Annually | 103 | 74.1 |
| Half-yearly | 33 | 23.74 |
| Monthly | 3 | 2.16 |
| Weekly | 0 | 0 |
| Total | 139 | 100 |

**6.6 Respondents Enrolled course in SWAYAM platform**

The distribution of the respondents enrolled courses their choice of subject which display in Table 7. Out of 186 respondents, 83.33% of respondents are enrolled course and 16.67% of the respondents of them not enrolled in any courses.

**Table 7: Respondent's enrolled course in SWAYAM platform**

| SWAYAM Registered | Respondents | Percentage |
| --- | --- | --- |
| Yes | 155 | 83.33 |
| No | 31 | 16.67 |
| Total | 186 | 100 |

**6.7 Respondents attending the mock test and submitting assignment every week**

Table 8 indicates enrolled courses to submit a mock test and assignment every week. The analysis reveals that the more than half of the 62.58% respondents are not attending the mock test and submitting assignments on the date and only 37.42% of them attended and submitted on the date.

**Table 8: Respondent's opinion for attending mock test & submitting an assignment on the date**

| Descriptions | Respondents | Percentage |
| --- | --- | --- |
| Yes | 58 | 37.42 |
| No | 97 | 62.58 |
| Total | 155 | 100 |

**6.8 Reason for not attending the mock test & submitting an assignment on the due date**

When the respondents were questioned about their reason for not attending the mock test and submitting an assignment on due date displayed in table 9. The analysis found that the majority 43.3% of the respondents are mentioned their lack of time, followed by 32.99% of them indicated heavy class routine. About 12.37% of them are due to system problems, 7.22% of them gave not support from the university and 4.12% are mentioned other reasons.

**Table 9: Reasons for not attending the mock test and submitting an assignment on the due date**

| Reasons | Respondents | Percentage |
|---|---|---|
| Lack of time | 42 | 43.3 |
| No support from the University | 7 | 7.22 |
| Heavy class routine | 32 | 32.99 |
| Due to systems | 12 | 12.37 |
| Other | 4 | 4.12 |
| Total | 97 | 100 |

**6.9 For Assignment submission documentformat used by the respondents**

Table 10 indicates the documents format preference for assignment submission. About 50.32% of the respondents used word file for assignment submission followed by PDF 27.1% and texting 18.71%. Only 3.87%t of them preferred PPT for assignment submission.

**Table 10 Format of assignment submission**

| Formats | Respondents | Percentage |
|---|---|---|
| Word file | 78 | 50.32 |
| PDF | 42 | 27.1 |
| Texting | 29 | 18.71 |
| PPT | 6 | 3.87 |
| Total | 155 | 100 |

**6.10 Electronic Devices Used for submitting an assignment**

Table 11 analysis shows that 52.26% of the respondents are used Laptop for submitting assignment followed by Desktop 34.19% and only 13.55% of them used Smartphone.

**Table 11: Electronic Devices Used for submitting an assignment**

| Device | Respondents | Percentage |
|---|---|---|
| Laptop | 81 | 52.26 |
| Desktop | 53 | 34.19 |
| Smartphone | 21 | 13.55 |
| Total | 155 | 100 |

**6.11 Types of resources referred for submitting an assignment**

The analysis shows that 76.13% of the respondents referred open access resources for assignment submission and 58.71% of them referred to books. It is followed by 52.26% referred to journal articles whereas 44.51% are E-Database.

**Table 12: Types of resources referred for submitting an assignment**

| Sources | Respondents | Percentage |
|---|---|---|
| Open Access Resources | 118 | 76.13 |
| Books | 91 | 58.71 |
| Journal Articles | 81 | 52.26 |
| E- Databases | 69 | 44.51 |

(Respondents are permitted to provide more than one answer)

## 6.12 University providing lab facility for SWAYAM course

Table 13 indicates the response to University providing lab facility for appearing SWAYAM courses. The analysis shows that 72.9% replied their University providing lab facility for SWAYAM courses remaining 27.1% replied negatively.

**Table 13 University provided lab facility for SWAYAM course**

| Lab facility | Respondents | Percentage |
|---|---|---|
| Yes | 113 | 72.9 |
| No | 42 | 27.1 |
| Total | 155 | 100 |

## 6.13 Respondents the Number of Courses Enrolled

Table 14 shows that highest 92.26% of the respondents enrolled 1-3 courses whereas only 5.94% enrolled in 4-6 courses. None of them is enrolled in 7-9 courses.

**Table 14: Respondents the number of courses enrolled**

| No of courses | Respondents | Percentage |
|---|---|---|
| 1-3 | 143 | 92.26 |
| 4-6 | 12 | 5.94 |
| 7-9 | 0 | 0 |
| Total | 155 | 100 |

## 6.14 Respondents of average time spending on the courses

Table 15 indicates the amount of time spent on an average for study courses enrolled by the students in various Universities. It is found that out of 155 respondents 82.58% were spent 1 – 3 hours for study course related materials and 12.9% spent 3 - 5 hours. About 4.52% of the respondents spent 5 - 7 hours and none of them are spending more than 8 hours.

**Table 15 Average time spending for the course**

| Time | Respondents | Percentage |
|---|---|---|
| 1 to 3 hours | 128 | 82.58 |
| 3 to 5 hours | 20 | 12.9 |
| 5 to 7 hours | 7 | 4.52 |
| > 8 hours | 0 | 0 |
| Total | 155 | 100 |

**6.15 Respondents of satisfaction level of course instruction and syllabus**

Table 16 shows the satisfaction of course instruction and syllabus among the various University students. The analysis found that the majority 55.48% of the respondents are satisfied with the course instruction and syllabus. About 25.17% of them neutral, 15.48% are highly satisfied and only 3.87% are not satisfied.

**Table 16 Satisfaction level of course instruction and syllabus**

| Level of Satisfaction | Respondents | Percentage |
|---|---|---|
| Highly Satisfied | 24 | 15.48 |
| Satisfied | 86 | 55.48 |
| Neutral | 39 | 25.17 |
| Not Satisfied | 6 | 3.87 |
| Total | 155 | 100 |

**6.16 Satisfaction level of SWAYAM Teaching Methods**

Table 17 indicates the satisfaction of SWAYAM teaching methods by the students of various Universities. It is revealed that most of the respondents 47.1% were highly satisfied with a video lecture, 38.06% of them satisfied, 13.55% are neutral and 1.29% are not satisfied. About 63.23% of the respondents are highly satisfied with reading materials, 26.45% were satisfied, 10.32% are neutral. Most of the respondents 56.13% of them highly satisfied with the mock test, 24.52% are satisfied, 18.06% of the neutral and only 1.29% are not satisfied with the mock test. Out of 155 respondents, 27.1% are highly satisfied with the assignment, 37.42% are satisfied and 35.48% are neutral.

**Table 17 Satisfaction level of SWAYAM Teaching Methods**

| Teaching Methods | Highly Satisfied | Satisfied | Neutral | Not Satisfied |
|---|---|---|---|---|
| Videos lecture | 73(47.1) | 59 (38.06) | 21 (13.55) | 2 (1.29) |
| Reading Materials | 98 (63.23) | 41 (26.45) | 16(10.32) | 0 |
| Mock Test | 87 (56.13) | 38 (24.52) | 28 (18.06) | 2 (1.29) |
| Assignment | 42 (27.1) | 58 (37.42) | 55 (35.48) | 0 |

**6.17 SWAYAM courses help gain new knowledge and support life-long learning**

Table 18 shows that majority 50.97% of the respondents are agreed SWAYAM courses help to gain new knowledge and supports life-long learning followed by 30.97% of them strongly agree and 9.68% of them neutral. About 4.52% are disagreeing and only 3.86% per cent of them strongly disagree.

**Table 18: Opinion on SWAYAM course**

| Opinion level | Respondents | Percentage |
|---|---|---|
| Strongly Agree | 48 | 30.97 |
| Agree | 79 | 50.97 |
| Neutral | 15 | 9.68 |
| Disagree | 7 | 4.52 |
| Strongly Disagree | 6 | 3.86 |
| Total | 155 | 100 |

**6.18 Opinion about the SWAYAM courses are similar to the regular class syllabus**

The analysis reveals that 40% of the respondents are neutral regarding the similarity of the regular class syllabus, 30.32% are agreed and 14.84% are strongly agreed. Only 11.61% of them disagree and 3.23% are strongly disagreed.

**Table 19: SWAYAM courses are similar to the regular class syllabus**

| Opinion level | Respondents | Percentage |
|---|---|---|
| Strongly Agree | 23 | 14.84 |
| Agree | 47 | 30.32 |
| Neutral | 62 | 40 |
| Disagree | 18 | 11.61 |
| Strongly Disagree | 5 | 3.23 |
| Total | 155 | 100 |

**6.19 Problem faced by the respondents for accessing SWAYAM courses**

Table 20 points out the difficulties faced by the respondents while accessing the SWAYAM courses. The analysis found that 50.32% replied routine classes there is no time to access, 20.65% faced slow internet and 12.26% found it difficult web architecture, 10.96% unable to access and only 5.81% faced other problems.

**Table 20: Difficulties to appear SWAYAM courses**

| Difficulties | Respondents | Percentage |
|---|---|---|
| Unable to access | 17 | 10.96 |
| Difficult web architecture | 19 | 12.26 |
| Slow internet | 32 | 20.65 |
| Routine classes | 78 | 50.32 |
| Others | 9 | 5.81 |
| Total | 155 | 100 |

**6.20 Number of persons withdrawing courses**

Table 21 shows the highest 78.06% of the respondents are continued the enrolled courses, whereas 21.94% of them are discontinued their courses.

**Table 21: Number of persons withdrawing courses**

| Withdraw courses | Respondents | Percentage |
|---|---|---|
| Yes | 34 | 21.94 |
| No | 121 | 78.06 |
| Total | 155 | 100 |

**6.21 Respondents withdrawn courses during their course period**

Table 22 indicates the number of respondent's with drawn courses during their course period. Majority 38.24% of the respondents have withdrawn during 5 to 6weeks, 32.35% of them during 3 to 4 weeks, 23.53% have withdrawn 1 to 2 weeks and only 5.88% withdrawn more than 6 weeks.

**Table 22: Number of persons during their course period withdrawing courses**

| During Withdraw courses | Respondents | Percentage |
|---|---|---|
| 1-2 weeks | 8 | 23.53 |
| 3-4 weeks | 11 | 32.35 |
| 5-6 weeks | 13 | 38.24 |
| >6 weeks | 2 | 5.88 |
| Total | 34 | 100 |

**6.22 Reasons for withdrawing courses**

Table 23 shows the highest 52.94% of the respondents are withdrawn SWAYAM courses due to no time, 17.65% of them thought that not very useful and equal number of respondents replied no assistance from the university. About 11.76% are having some other problems.

Table 23 Reasons for withdrawing courses

| Reasons | Respondents | Percentage |
|---|---|---|
| Not very useful | 6 | 17.65 |
| No time | 18 | 52.94 |
| No assistance from the university | 6 | 17.65 |
| Other | 4 | 11.76 |
| Total | 34 | 100 |

7. **Findings**

Most of the responses 41.9% received from Annamalai University followed by 13.97% responses received from the Central University of Tamil Nadu. Majority of the respondents 87.63% belong to the age group of 18-24 years followed by 9.13% of the respondents belong to the age group of 25-30 years. In this survey 67.74% of the respondents are Male and the remaining 32.26% are female. Majority of the respondents 83.33% are aware of SWAYAM courses remaining 16.67% are unaware. Most of the respondents 41.29% known about the SWAYAM course through their teachers and 22.58% from friends. In response to orientation programme providing by the university, 74.73% of respondents replied positively. Most of the respondents 62.58% are not attending the mock test and submitting assignments on date and remaining 37.42% of them are attending and submitting on the date. In response to reasons for not attending the mock test, 43.3% of the respondents are mention lack of time, followed by 32.99% of them indicated routine classes. Half of the respondents 50.32% are using word file for assignment submission followed by PDF 27.1%. More than half of the respondents 52.26% are using Laptop for submitting assignment followed by 34.19% Desktop.

For assignment submission 76.13% respondents referring open access resources, more than half of the respondents referring books and journals and 44.51% are referring E-Databases.

72.9% replied their University providing lab facility for SWAYAM courses access remaining 27.1% replied negatively. Highest 92.26% of the respondents enrolled 1-3 courses in SWAYAM whereas only 5.94% enrolled in 4-6 courses. 82.58% of respondents were spending 1 – 3 hours for study course related materials and 12.9% spent 3 - 5 hours. In response to the satisfaction level of SWAYAM course, 55.48% of the respondents are satisfied with the course instruction and syllabus, 15.48% are highly satisfied. 47.1% were highly satisfied with video lecture; about 63.23% of the respondents are highly satisfied with reading materials. Most 56.13% of them highly satisfied with mock test. Out of 155 respondents, 27.1% are highly satisfied with the assignment, 37.42% are satisfied and 35.48% are neutral.

Half of the respondents 50.97% are agreed SWAYAM courses help to gain new knowledge and supports life-long learning followed by 30.97% of them strongly agreed. In response to an opinion about the SWAYAM courses are similar to the regular class syllabus, 40% of the respondents are neutral, 30.32% are agreed and 14.84% are strongly agreed. In response to the problem faced by the respondents while accessing SWAYAM course analysis found that 50.32% replied heavy class there is no time to access, 20.65% faced slow internet. It is found that 78.06% of respondents are continuing their enrolled courses remaining are withdrawn. In response to the reason for withdrawing SWAYM courses, 52.94% of the respondents are replied no time, 17.65% of them thought that not very useful and no assistance from the University.

8. **Conclusion**

SWAYAM is developing for the welfare of the young minds that has the right to access the learning sources and to enrich their skills in the required field with no cost. The researcher has found from the survey taken on the programme SWAYAM has its positive way of producing and bringing out young talented mind to society. The result shows us the respondents have enthusiasm towards acquiring the knowledge through online courses because it saves their time and it teaches us at any time. The learning resources can be viewed many times until the respondents understand the subject. The result gives us the report that male respondents are more involved in doing the online courses where female respondents are not interested to do online. The SWAYAM course has its disadvantage in dealing with test and assignments. Though the courses work with no cost it is clearly shown that the courses were discontinued if the learners were not monitored. Many of the courses have been discontinued by the learners in the middle of the session because of their lethargic behaviour. The learners prefer to complete courses which provides them with certificates. The survey reports us the learners were aware of SWAYAM programme which has a motto towards providing new knowledge and supports life-long learning. The report clearly states that when the programme is also introduced in other field students, the students involve themselves to enrich their knowledge with the support of the SWAYAM course.